\documentclass[11pt,twoside]{article}

\usepackage{asp2004}
\usepackage{epsf}
\usepackage{psfig}
\usepackage{lscape}
\usepackage{graphics}

\begin{document}
\title{Blue Stragglers, Be stars and X-ray binaries in open clusters}   
\author{Amparo Marco, Ignacio Negueruela}  
\affil{DFISTS, EPSA, Universidad de Alicante, E-03080,
  Alicante, Spain}
\author{Christian Motch}
\affil{Obs. de Strasbourg, 11 rue de l'Universit\'{e}, F67000,
Strasbourg, France} 

\begin{abstract} 
Combination of high-precision photometry and spectroscopy allows the detailed 
study of the upper main sequence in open clusters. We are carrying out
a comprehensive study of a number of clusters containing Be stars in
order to evaluate the likelihood that a significant number of Be stars
form through mass exchange in a binary. Our first results show
that most young open clusters contain blue stragglers. In spite of the
small number of clusters so far analysed, some trends are beginning
to emerge. In younger open clusters, such as NGC~869 and NGC~663,
there are many blue stragglers, most of which are not  
Be stars. In older clusters, such as IC~4725, the fraction of Be stars among 
blue stragglers is very high. Two Be blue stragglers are moderately
strong X-ray sources, one of them being a confirmed 
X-ray binary. Such objects must have formed through binary
evolution. We discuss the contribution of mass transfer in a close
binary to the formation of both blue stragglers and Be stars 


\end{abstract}


\section{Introduction}   
Blue stragglers (BSs) are stars lying above the main sequence (MS) turnoff
region in colour-magnitude diagrams, a region where, if the BSs had
been normal single stars, they should already have evolved away from
the main sequence \citep{stry}. Several mechanisms have been proposed to
explain the formation of BSs in different environments. 
\begin{itemize}
\item BSs may be stars formed later than the rest of the cluster or
  association. Though this may well happen in some regions with
  sequential star formation, in many clusters there are no stellar
  sequences connecting the BSs with the turn-off, arguing against a
  second epoch of star formation.
\item BSs might be evolved stars back in the blue region of the HR diagram
  after a red supergiant phase. However, abundances of Nitrogen in
  blue stragglers are much lower than those predicted by models for
  stars in blue loops \citep{smartt}.  
\item BSs may be stars that, for some reason, have evolved
  bluewards. In particular {\it homogeneous evolution} has been
  proposed as a mechanism that can result in blueward evolution for
  very high (near-critical) initial rotational velocity
  \citep{maeder}. However there 
  is no strong {\it a priori} reason to believe that many such extreme
  rotators will be formed. 
\item BSs may be formed by coalescence of two stars. This mechanism
  requires a high stellar density environment. Though it is probably the
  major channel for the formation of BSs in globular cluster, it
  is unlikely to be able to explain BSs in OB associations.
\item BSs result from mass exchange in close binaries. Examples of this
  process abound amongst massive binaries (see review by Negueruela in
  these proceedings). Therefore this channel must contribute some BSs in
  young clusters and associations.
\end{itemize} 

While the properties of BSs in globular clusters have been the object
of many studies, the mechanisms for forming BSs in young open clusters
have deserved little attention. The most relevant works are those of
\citet{mermi} and \citet{mathys} 

\section{Observations}

As part of our programme to fully characterise a sample of young open
clusters, we have obtained spectroscopy of large numbers of stars in
the young open clusters NGC~663, NGC~869 and NGC~3766, and the
moderately older clusters NGC~6649 and IC~4725.  

We obtained spectra for 38 bright stars in $h$ Persei and the
immediately surrounding area on 2003 November 15th, using the 4.2-m
WHT at the La Palma Observatory, equipped with the ISIS spectrograph
and the R600B grating (0.4 \AA/pixel). In addition, 17 members were
observed with the 2.5-m INT, also at the La Palma Observatory, with the
IDS spectrograph and the R900V grating (0.6 \AA/pixel), in July 2003.

Spectra for 140 members of NGC 663 were obtained in two runs during
October/November 2002, one with the INT+IDS and the R400V grating (1.2
\AA/pixel) and the other one with the 1.93-m telescope at Observatoire
de Haute-Provence and the Car\'elec spectrograph and the 300 ln/mm
grating (1.8 \AA/pixel). 

Spectra of the brightest members of NGC 6649 were obtained with the WHT+ISIS
and the R300B grating (0.9 \AA/pixel) on 2004 May 20th. 

Finally observations of NGC 3766 and IC 4725 were obtained in May 2004 with
the New Technology Telescope (NTT) at La Silla, equipped with EMMI in
the longslit spectrograph mode.


\section{Results}

\subsection{HR diagrams}
In order to detect BSs in our sample of open clusters, we use two
types of HR diagram. When Str\"omgren photometry of a sufficient large
sample exists, we use the observational $M_{V}/c_{0}$ diagram for the
determination of cluster ages. If we have spectroscopy of a large
number of members, we can also construct a ``theoretical'' HR diagram
using the following procedure: from our spectra, we determine accurate
spectral types and then take the corresponding $T_{{\rm eff}}$  from
the tabulation of \citet{hme84}. We then calculate the individual
reddening for each star, using precision $UBV$ (and $RI$ where
available) photometry from the literature and $JHK$ from 2MASS. The
photometry and the Kurucz model corresponding to the spectral type are
used as input for the $\chi^{2}$
code for parameterised modelling and characterisation of photometry and
spectroscopy implemented by \citet{maiz04}, which estimates the
reddening . After dereddening, we 
correct for the adopted distance modulus and transform to $M_{{\rm
    bol}}$ using the calibration of BC for a given spectral type
given by \citet{hme84}. All the isochrones shown are from
\citet{schaller}. 
 
\subsection{NGC 3766}
\label{sec:3766}

\begin{figure}[t!]
\label{fig:phot3766}
\plotone{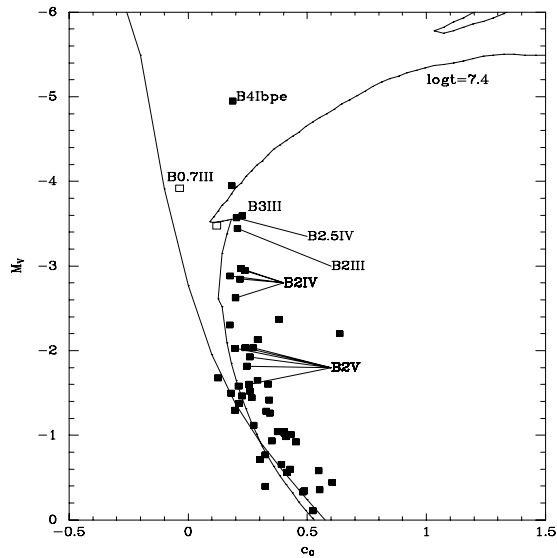}
\caption{$M_{V}-c_{0}$ diagram for NGC 3766, based on the photometric
  data from \citet{shob85} and \citet{shob87}, with the spectral types
  derived from our spectra.} 
\end{figure}

We have Str\"{o}mgren photometry for 55 stars in this cluster from
\citet{shob85} and \citet{shob87}. The cluster HR digram
(Fig.~1) clearly demonstrates the age of the cluster,
$\sim25\:$Myr. This age is fully consistent with the spectral types of
the stars around the MS turn-off (B2\,V). However, there is an object
with $M_{{\rm V}} = -3.9$ lying clearly to the left of the
isochrone. This star (S5) is located
at the very centre of the cluster. Its spectral type B0.7\,III
confirms that it is earlier than any other member, but also that its
parameters (distance and reddening) are those of a cluster member. We
thus conclude that this object is a blue straggler in NGC~3766.  

The other object marked with an open square
(S326, BF~Cen) also is a very interesting case. Its spectrum 
clearly shows that it is a
double-line spectroscopic binary. \citet{helt} found that it is an
eclipsing binary with a 3.7-d orbital period. The primary component is
a B1.5\,V star with 8.7$M_{\sun}$ and 5$R_{\sun}$, while the
secondary component is a B6\,III star with 3.8$M_{\sun}$ and
7$R_{\sun}$. The latter rotates 
synchronously, while the primary rotates much faster, implying that it
is a post-mass transfer system. Taking into account that the secondary
contributes a substantial amount of light, the B1.5\,V star alone
would be located to the
left of the turn-off and be a mild blue straggler. This is a good
example of a blue straggler created by mass transfer. 

\subsection{NGC 869}

 In the observational photometric diagram of NGC~869,
 there are many stars to the left of the MS turn-off
 \citep{marco}. This is also evident in the theoretical HR
 diagram (Fig.~2) built from our spectral types and the photometry
 from \citet{kel01}.  The spectral types of the stars leaving the main
 sequence (B1.5\,V) are compatible with an age of
 $\sim15-20\:$Myr. However, a large number of evolved stars lying to
 the left of the turn-off have spectral types indicating higher
 masses.

\begin{figure}[t]
\label{fig:phot869}
\plotfiddle {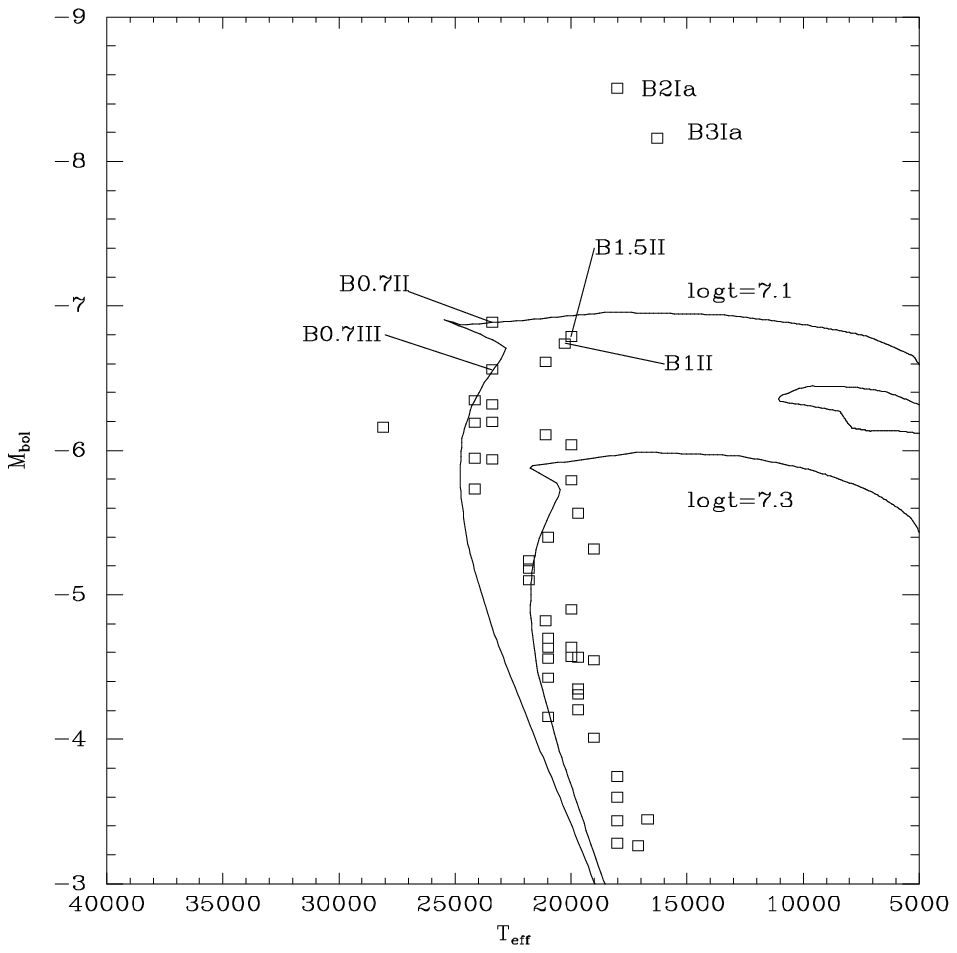}{3cm}{0}{80}{80}{-370}{-320} 
\plotfiddle {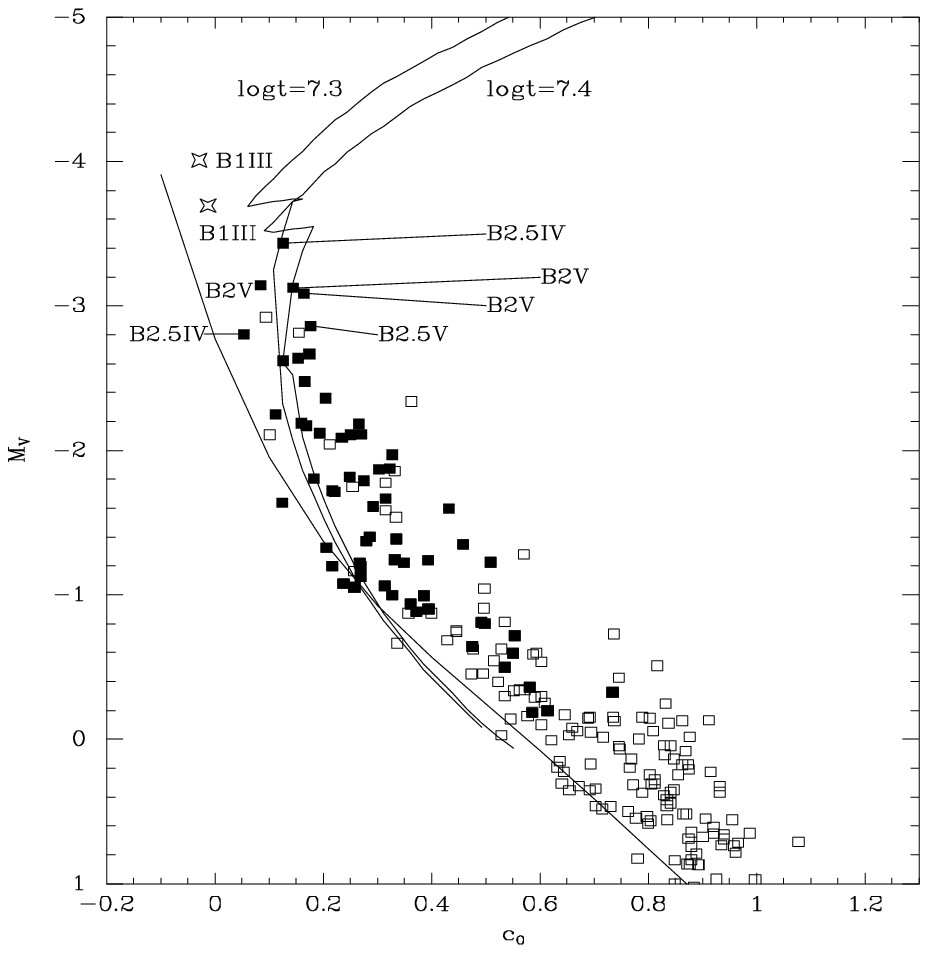}{3cm}{0}{80}{80}{-120}{-223}
\caption{{\bf Left panel} Theoretical HR diagram for h Persei built
from the observations following the procedure indicated in the
text.\newline
{\bf Right panel} $M_{V}-c_{0}$ diagram for NGC 663. The stars marked
with filled squares were observed spectroscopically. Spectral types
for some members are shown. The objects marked with stars are spectroscopic and photometric blue stragglers.}
\end{figure}

We can distinguish two kinds of BSs.  Most of them concentrate around
the $\log t=7.1$ isochrone. Though
\citet{strom} show that many of these objects are not very fast
rotators {\it now}, the fact that all of them are evolved from the MS
leads us to speculate that their position to the
left of the turn-off may be related to initial fast rotation. 
A few other BSs occupy positions indicating much higher masses. The
B2\,Ia and B3\,Ia supergiants are amongst the most massive B-type
supergiants in the sample of \citet{mcerlean}, with evolutionary
masses of the order of $\sim30M_{\sun}$. Both types of BSs are present
at the cluster core and most of them are certainly cluster members.  

\begin{figure}[t]
\label{fig:two}
\plotfiddle {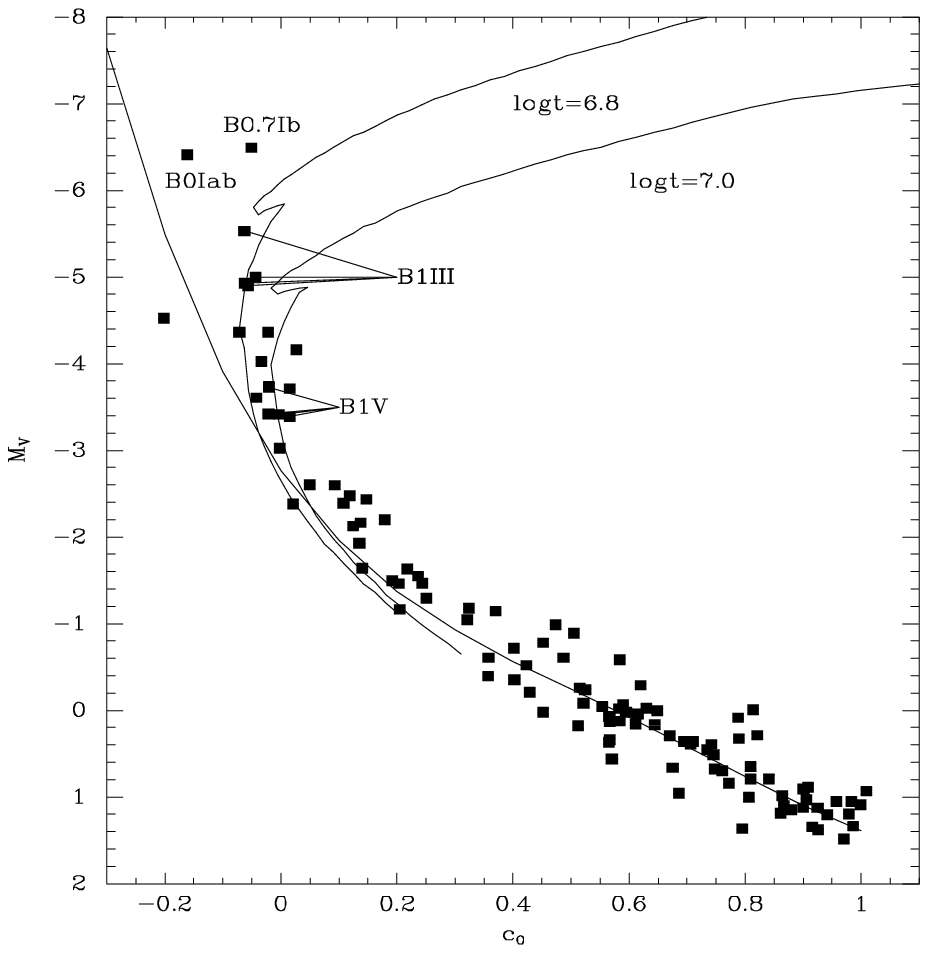}{3cm}{0}{80}{80}{-370}{-320} 
\plotfiddle {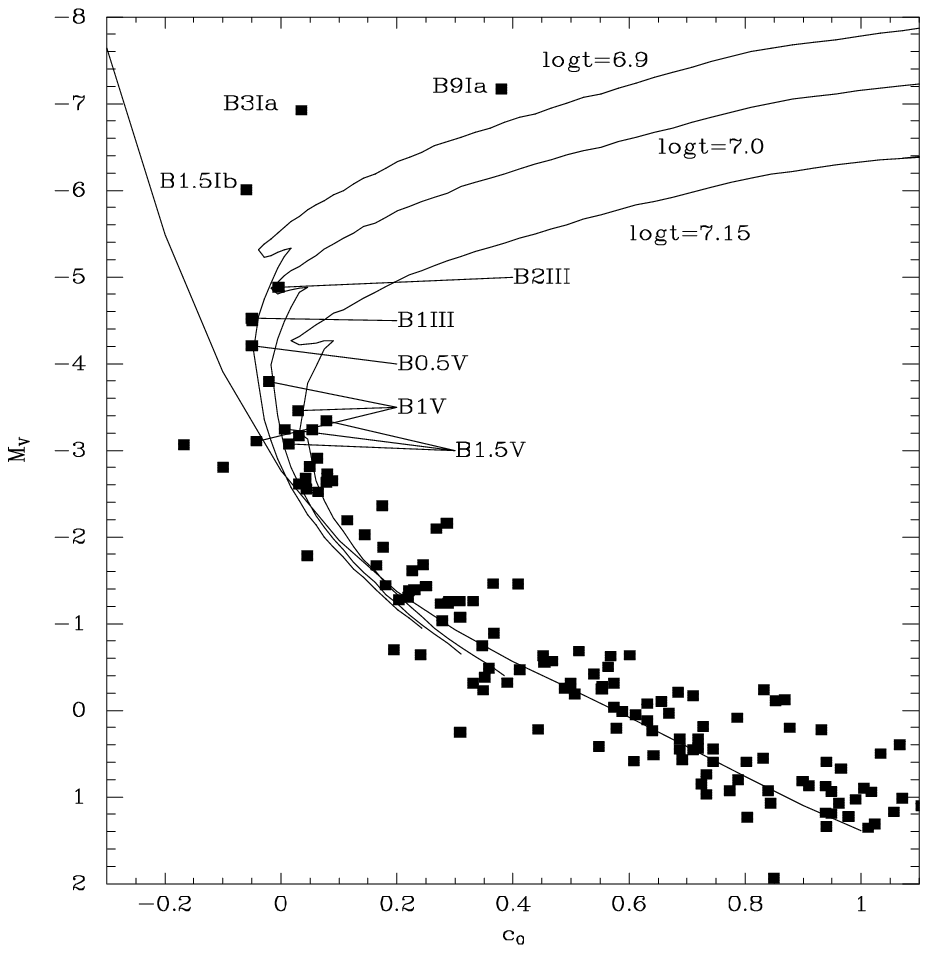}{3cm}{0}{80}{80}{-120}{-223}
\caption{{\bf Left panel} $M_{V}-c_{0}$ diagram for NGC 3293, based on
  the photometric data from \citet{balona94}, with spectral types from
  \citet{evans2005} 
\newline
{\bf Right panel} $M_{V}-c_{0}$ diagram for NGC 4755, based on the
photometric data from \citet{koen94}, with spectral types from
\citet{evans2005}.} 
\end{figure}

A similar effect, well-defined MS turn-off and most of the evolved
stars lying to the left, can also be seen in two other relatively massive
clusters, perhaps slightly younger than NGC~869, for which abundant
data exist in the literature: NGC 3293 and NGC 4755 (see
Fig.~3).  
In these three open clusters and many others of similar age, there are
B-type supergiants lying well above the corresponding isochrones for
the cluster. Their progenitors are stars more massive than
$\sim20M_{\sun}$ and therefore they are necessarily BSs in clusters
older than $10\:$Myr.

\subsection{NGC 663}

From the observational photometric diagram, the cluster has an age
$\sim25\:$Myr, again in good agreement with the spectral types of
stars close to the turn-off (B2\,V). We find two obvious photometric
BSs located near the cluster core (S4 and S30). Their spectral 
types (B1\,III in both cases) support their BS nature and their
cluster membership. Our spectra reveal at least three other
spectroscopic BSs which were saturated in our
photometry. They include the O9.5\,V star S162 and, of particular interest,
LS I +61$^{\circ}$ 235 (S194), a B0.5\,IVe star in the Be/X-ray binary
RX~J0146.9+6121.

\subsection{NGC 6649}

The age of this cluster is $\sim50\;$Myr \citep{wl87}, confirmed by the
study of the characteristics of the double mode Cepheid member V367
Sct. We build the theoretical HR diagram from our spectra and the
photometry of \citet{wl87}. The spectral types of the brightest blue members  
show good agreement with this age, presenting a MS turn-off around B4\,V 
(Fig.~4). There are two obvious BSs, the brightest of which is
a B1\,IIIe star (S9), unambiguously associated with an {\it XMM-Newton}
X-ray source. This 
object appears to be a low luminosity Be/X-ray binary or a member of
the class of $\gamma$-Cas-like X-ray sources (see contribution by
Motch et al. in these proceedings).

\subsection{IC 4725}
The age of this cluster has been set at $90-100\:$Myr by several
different studies and is also constrained by the presence of a classical
Cepheid, U Sgr. From the spectral types, we find the MS turn-off
around B6\,V, in  good agreement with this age, but leaving at least
four evolved BSs, three of which are Be stars (see Fig.~4).   

\begin{figure}[t!]
\label{fig:dia6649}
\plotfiddle {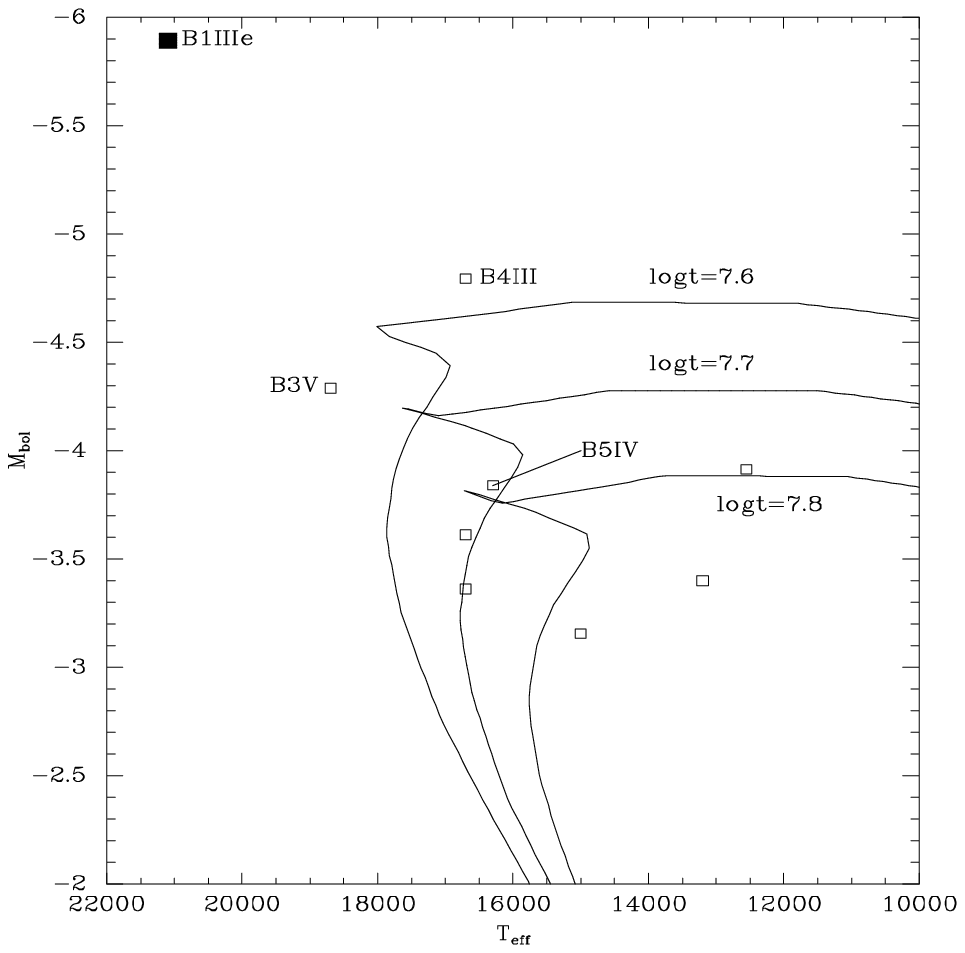}{3cm}{0}{80}{80}{-370}{-320} 
\plotfiddle {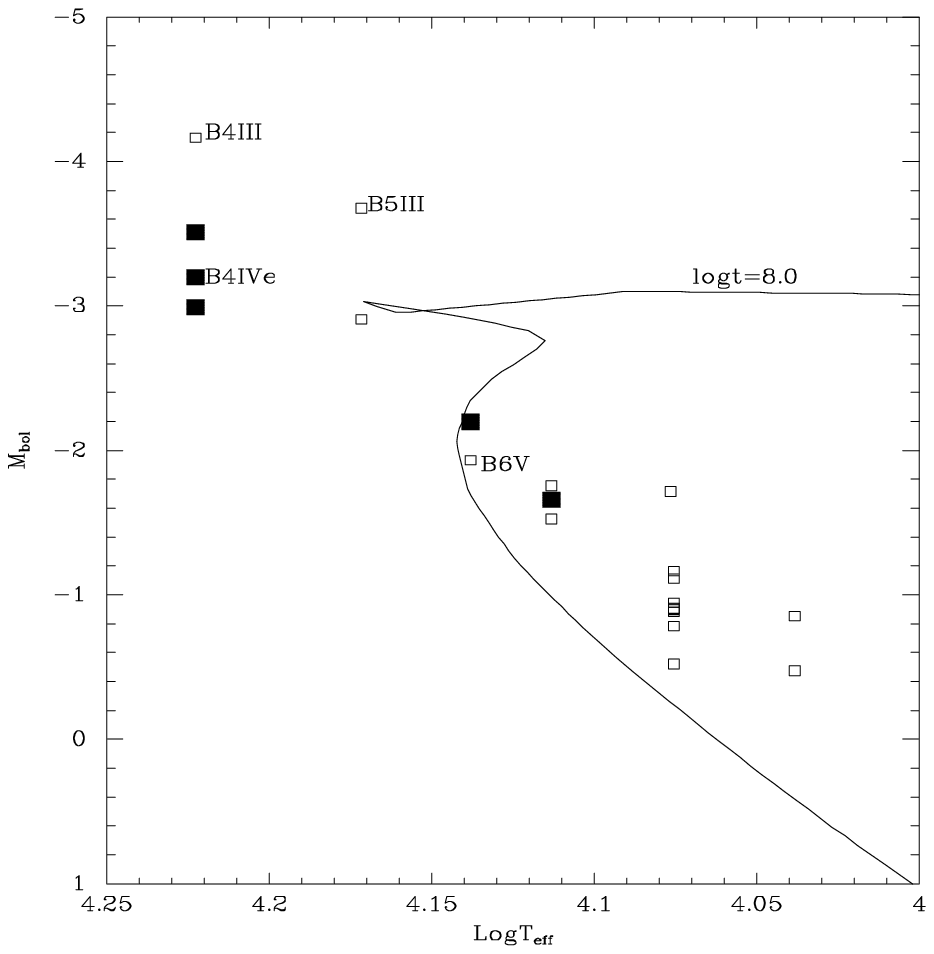}{3cm}{0}{80}{80}{-120}{-223}
\caption{{\bf Left panel} Theoretical HR diagram for NGC 6649 built
from the observations following the procedure indicated in the
text.\newline
{\bf Right panel} Theoretical HR diagram for IC 4725 built
from our spectra and photometry by \citet{joh60}. Be stars appear as
filled squares.}
\end{figure}
  
\section{Discussion and conclusions}
Though the sample of open clusters for which a detailed analysis has
been made is still quite small, we observe the following interesting trends:

\begin{itemize}
\item Blue stragglers are found in all open clusters surveyed
\item In clusters $\sim15\:$Myr old, most evolved stars appear too
  massive for these ages. 
In particular, many of these clusters  contain Ia and Iab supergiants,
  which must be descended from stars with $\ga20M_{\sun}$ according to all
  theoretical evolutionary paths. Such massive stars have lifetimes
  $<10\:$Myr \citep{mm03}, rendering all these supergiants BSs.
\item Among clusters younger than $\sim25\:$Myr, very few BSs are Be stars.
\item From the data on IC~4725 and NGC~6649, incomplete samples in
  NGC~2516 and IC~2488 and data in the literature \citep{mermi}, it
  would seem that most BSs in clusters in the $50-150$~Myr range are
  Be stars. 
\item Our data suggest that a non-negligible fraction of BSs in moderately
young open clusters are formed by mass transfer in a close binary
(X-ray sources in NGC~663 and NGC~6649, eclipsing binary in NGC~3766),
but not all BSs are binaries, making it unlikely that this might be the only
channel. 
\item Likely different mechanisms are dominant at different ages.
\end{itemize}

\acknowledgements 
This research is partially supported by the Spanish
  MCyT under grant AYA2002-00814 and the Generalitat Valenciana under
  grant GV04B/729. IN is a researcher of the
programme {\em Ram\'on y Cajal}, funded by the Spanish Ministerio de
Ciencia y Tecnolog\'{\i}a and the University of Alicante.

The INT and WHT are operated on the island of La
Palma by the Isaac Newton Group in the Spanish Observatorio del Roque
de Los Muchachos of the Instituto de Astrof\'{\i}sica de Canarias. Some
observations presented here were obtained as part of the ING service
programme. Partially based on
    observations collected at the European Southern Observatory,
    Chile (ESO 73.D-0032). Partially based on 
observations made at 
Observatoire de Haute Provence (CNRS), France.
This research has made use of the Simbad data base, operated at CDS,
Strasbourg (France) and the WEBDA open cluster database.


\end{document}